%% file: lp99.tex
\documentstyle[12pt,epsfig]{article}

\input LP99macros.tex

\def\Title#1{\begin{center} {\Large #1 } \end{center}}

\begin{document}

\Title{Lattice Calculations and Hadron Physics}

\bigskip\bigskip


\begin{raggedright}  

{\it S. Aoki\index{Aoki, S.}\\
Institute of Physics \\
University of Tsukuba,
Tsukuba, Ibaraki 305-8571, Japan }
\bigskip\bigskip
\end{raggedright}

\section{Introduction}

Lattice QCD \index{lattice QCD} 
aims to understand the strong interaction of hadrons from the first principles
of QCD for quarks \index{quarks} and gluons \index{gluons} with the 
aid of numerical simulations.  The physical quantities calculated 
with the method are many, ranging from the spectrum of light hadrons 
to a variety of electroweak matrix elements.   
Practical limitations such as finite lattice volume and spacing, and 
the use of the quenched approximation \index{quenched approximation} of 
ignoring sea quarks are gradually being lifted due to development of computer 
power, particularly those of dedicated parallel computers.  
In this talk we review progress in lattice QCD,
focusing on efforts to calculate weak matrix 
elements \index{weak matrix elements} relevant
for the determination of the Cabibbo-Kobayashi-Maskawa (CKM) matrix
\index{CKM matrix}.

We start the review with discussion of the latest results for the 
light hadron spectrum \index{light hadron spectrum} and 
quark masses \index{quark mass} in sect.~\ref{sec:light}. 
In addition to being a topic basic to all of lattice QCD, 
the recent progress in the quenched and full QCD \index{full QCD} calculations 
has yielded an interesting result, relevant for the CKM determination 
as well, that dynamical sea quarks have a significant effect 
on the value of the strange quark mass. \index{strange quark mass}

The CP-violation \index{CP-violation} in $K$ meson \index{K meson} decays 
continues to be an active topic with recent experimental reports on 
$\varepsilon'/\varepsilon$\cite{KTeV,Na48}.    
In sect.~\ref{sec:kcp} we discuss the present status of
the lattice calculation of the $K$ meson mixing parameter 
\index{mixing parameter} $B_K$ and the 
decay amplitudes including the $\varepsilon'/\varepsilon$ for which 
there has been recent renewed attempts. 

Understanding physics of $B$ mesons \index{B meson}
is very important for the determination 
of the CKM matrix in the Standard Model \index{Standard Model} 
and to detect the physics beyond it.
It will be even more so as detailed data from the $B$ factories, 
\index{B factories}
BaBar\cite{babar} and Belle\cite{belle}, become available.
In sect.~\ref{sec:heavy} we present the lattice effort for calculating 
the weak matrix elements of heavy-light mesons. \index{heavy-light mesons}
After a brief discussion on the theoretical issues of heavy quark calculations
on a lattice, we present recent results on the $B$ meson decay 
constant \index{decay constants} $f_B$,
the mixing parameter $B_B$ and the form factors \index{form factors} 
of semi-leptonic decays.
For the decay constant serious attempts exist already to detect sea quark 
effects in its value.  The other, more complicated quantities, are still 
calculated in the quenched approximation. 
For technical details and recent developments
of heavy quark physics on the lattice we refer to 
recent reviews at the lattice conferences\cite{lat99,lat98}. 

In sect.~\ref{sec:ckm} we sketch to what extent the lattice results help 
constrain the CKM matrix determination.  We conclude with a brief 
summary in sect.~\ref{sec:end}.

\section{Light hadron spectrum and quark masses}
\label{sec:light}
\subsection{Light hadron spectrum}
Calculation of the light hadron spectrum has been pursued since 1981
because an agreement of the calculated 
spectrum with experiment provides a fundamental confirmation for the 
validity of QCD in the non-perturbative low-energy domain, and also 
gives a measure of the reliability of the calculational techniques.

\begin{figure}[tbh]
\begin{center}
\epsfig{file=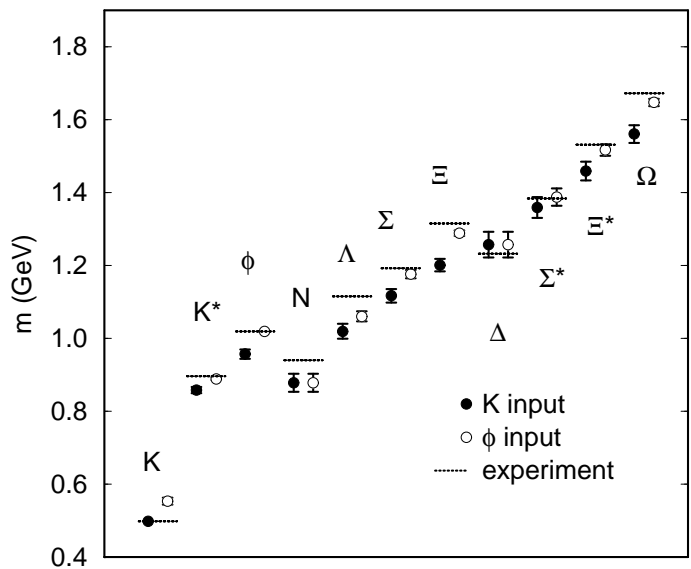, height = 7.5cm}
\caption{Light hadron spectra in Quenched QCD.}
\label{fig:spectrum}
\end{center}
\end{figure}

A recent precision result of the CP-PACS \index{CP-PACS} 
Collaboration\cite{qCP-PACS} for 
the light hadron spectra in the continuum limit \index{continuum limit} 
within the quenched approximation is shown in Fig.~\ref{fig:spectrum}. 
The up and down quarks are assumed to be degenerate ($m_u = m_d \equiv  m_l$)
in this calculation, and the experimental $\rho$ and neutral $\pi$ meson 
masses are used to fix the scale 
(lattice spacing $a$) and the light quark mass ($m_l$).  For 
the strange quark mass, two choices are compared, one employing 
the $K$ meson mass (filled symbols ; $K$-input) and other with 
the $\phi$ meson mass (open symbols; $\phi$-input). 
Experimental values are shown by horizontal lines.

This result shows an overall agreement of the light hadron spectrum 
in the quenched lattice QCD at a 5--10\% level, as previously demonstrated by
the GF11 \index{GF11} Collaboration\cite{GF11}.
However, it is also clear that there is systematic deviation between 
the quenched spectrum and experiment beyond the calculational error 
of 2--3\%. In particular, the hyperfine splitting \index{hyperfine splitting}
between the $\phi$, $K^*$ meson masses and the $K$ meson mass is smaller than
the experimental one.

\begin{figure}[tbh]
\begin{center}
\epsfig{file=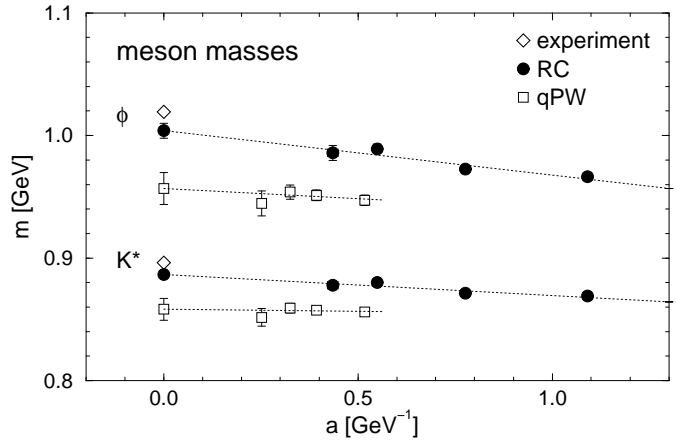, height = 6.0cm}
\caption{$\phi$ and $K^*$ meson masses from $K$ input
as a function of $a$ in full QCD (filled symbols), 
together with the quenched results (open symbols).}
\label{fig:meson}
\end{center}
\end{figure}

As is shown in Fig.~\ref{fig:meson}, however,
$K^*$ and $\phi$ meson masses from $K$ input 
agree much better in the 2 flavor full QCD\cite{fCP-PACS} than 
in the quenched QCD with the experimental values
after taking the continuum limit.
The effect of dynamical sea quarks is really
important for reproducing the correct spectra.

\subsection{Light quark masses}
The masses of quarks are fundamental parameters of the Standard Model.
Because of quark confinement, \index{confinement}
they can only be determined indirectly 
from a comparison of experimental hadron masses and their theoretical 
prediction in terms of quark mass parameters.  
For these reasons, much effort in lattice calculations have been devoted 
for their extraction. In this section we discuss the determination of 
strange quark mass $m_s$ which, as we shall see below, has an important 
impact on the $\varepsilon'/\varepsilon$ in $K$ meson decays. 

\begin{figure}[bht]
\begin{center}
\epsfig{file=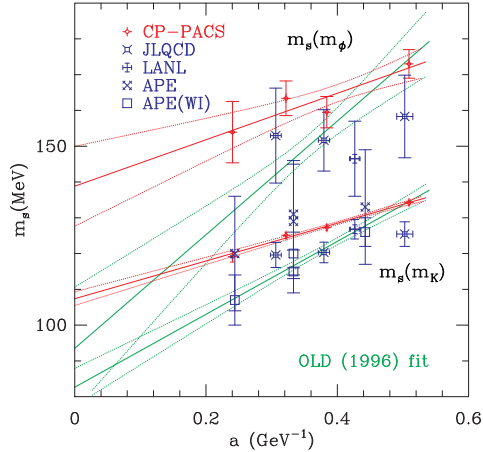, height = 6.0cm}
\caption{Strange quark mass in quenched QCD with Wilson quark actions.
Two continuum extrapolations are shown for each input:
one is the fit using all but the CP-PACS data, the other is the fit using
only the CP-PACS data.}
\label{fig:BG}
\end{center}
\end{figure}

A summary of results for $ m_s^{\overline{MS}}$(2 GeV) in quenched QCD 
with the Wilson quark action\cite{wilson}, 
compiled by Bhattacharya and Gupta in 
1997\cite{BG},
is shown in Fig.~\ref{fig:BG} as a function of $a$ together with 
linear continuum extrapolations as explained in the caption. 

From this analysis for results with the Wilson quark action and similar 
ones for the other types of quark actions, 
they concluded as their best estimate of $\overline{MS}$ mass at 2 GeV 
in quenched QCD to be
$
m_s^{\overline{MS}} ( 2 {\rm GeV}) = 110 (20) (11) {\rm MeV},
$
where the first error is estimated from the difference in the results among
Wilson, Clover\cite{sw} and KS\cite{ks} quark actions,
and the second one is the uncertainty of the scale ($1/a$) determination.
This value is already close to the lower bound
from QCD sum rules\cite{SumRule},
$m_s^{\overline MS}$(2 GeV) $\ge$ 90 - 100 MeV.

Two problems were manifest in their analysis. 
First, as can be seen from the figure, the large spread of data at finite
$a$ makes reliable continuum extrapolation difficult. 
This difficulty can be overcome by high precision calculations. 
Indeed the CP-PACS data in the figure is precise enough for
a reasonable linear extrapolation.

A more difficult problem, made quite clear by the CP-PACS precision data, 
is that the value of strange quark mass from $K$-input disagrees with 
the one from $\phi$-input, even in the continuum limit. 
This is one of the manifestation of the quenching error.

\begin{figure}[htb]
\begin{center}
\epsfig{file=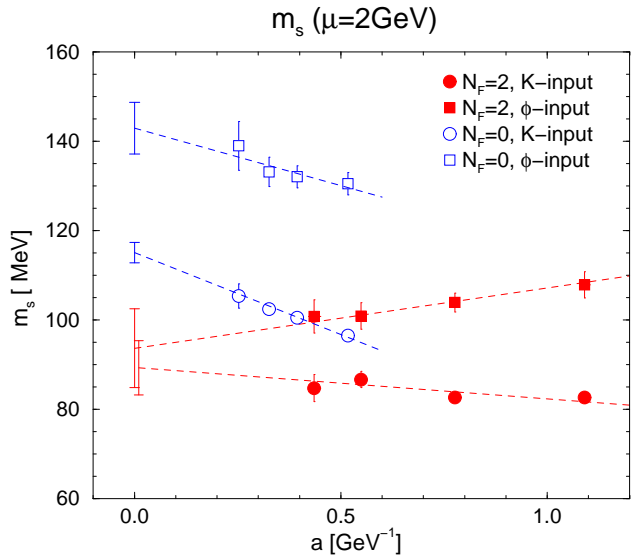, height = 7.5cm}
\caption{Strange quark mass in full QCD with Clover quark action
(filled symbols) as well as the one in quenched QCD with Wilson quark action
(open symbols), from $K$-input (circles) and $\phi$-input (squares).}
\label{fig:fCP-PACS}
\end{center}
\end{figure}

The full QCD result with 2 flavors of dynamical quarks
from the CP-PACS collaboration this year\cite{fCP-PACS} is shown in 
Fig.~\ref{fig:fCP-PACS}.
Here the strange quark mass 
$m_s^{\overline{MS}}$ (2GeV), defined through the axial-vector Ward-Takahashi
identity, \index{Ward-Takahashi identity} is plotted as a function of $a$.
Filled symbols in the figure are full QCD results\cite{fCP-PACS} obtained
with the Clover quark action and renormalization group (RG) improved gauge
action, while open ones are the previous quenched results with
the Wilson quark action and the ordinary plaquette gauge action\cite{qCP-PACS}.
Results from both $K$-input (circles) and $\phi$-input (squares) are
given as in the previous figure.
Numerically, we obtain in the continuum limit that 
$m_s^{\overline{MS}}$ (2 GeV) = 89.3 (6.6) MeV from
$K$-input and 93.7 (8.8) MeV from $\phi$-input in the 2 flavor full QCD,  
while 115 (2) MeV and 143 (6) MeV are the values for the 
$K$- and $\phi$-input, respectively, in quenched QCD.

It is very encouraging to observe that strange quark masses from two different
inputs agree with each other within errors in the continuum limit. 
This reflects the fact, discussed already, that 
the hyperfine splitting among stranged mesons agrees well
with experiment in the 2 flavor full QCD.

It is surprising, however, that the quenching errors on the strange quark mass
are larger than expected:
20\% for $K$-input and 40\% for $\phi$ input. Moreover the strange quark mass
in full QCD is much smaller than the previous quenched one, and is very close
to the lower bound from QCD sum rules\cite{SumRule}.

We quote the averaged value from the two inputs,
\beq 
m_s^{\overline{MS}} ({\rm 2 GeV}) = 91(13) {\rm MeV},
\eeqn
as the CP-PACS result for the strange quark mass at 2 GeV in 2 flavor full QCD,
which is equivalent to 
$m_s^{\overline{MS}} (m_c)$ = 100(15) MeV at the scale of the charm quark mass.

Let us comment on the systematic uncertainties in this result:
(i) The 2 flavor QCD simulation still neglects the dynamical effect of 
the strange quark itself.  Although this effect
is expected to be smaller than that from zero to 2 dynamical flavors, 
the value of the strange quark mass may be further reduced.
(ii) The lightest dynamical sea quark masses used in the simulation 
corresponds the $\pi$ over $\rho$ mass ratio of $m_\pi/m_\rho=0.55$.
One has to extrapolate the result obtained at these heavier quark masses
to the physical light quark mass, where  $m_\pi/m_\rho = $ 0.18.
Their may be systematic uncertainties associated with opening of decay 
channels such as $\rho\to 2\pi$ which are difficult to assess for 
the current range of heavy sea quarks. 
(iii) For the renormalization factor \index{renormalization factor}
to convert the bare quark mass on 
the lattice into the renormalized $\overline{MS}$ quark mass in the 
continuum, the perturbative value calculated to one-loop order 
is used.  
It is certainly desirable to use the renormalization factor
calculated non-perturbatively\cite{NPZ}.  For the present analysis, 
effects of higher order terms neglected are partly estimated by 
comparison of the quark mass evaluated through the axial-vector 
Ward-Takahashi identity and those from the mass parameter in the action.
The influence on $m_s^{\overline{MS}}$ (2 GeV) is found to be 
2-3 MeV, which has already been included in the estimate of the
central value and the error.
(iv) There exists also an uncertainty in the continuum extrapolation.
Variations of the strange quark mass 
by a change of the ansatz in the continuum extrapolation,
from linear in $a$ to, for example, linear plus quadratic in $a$ 
or $ a\cdot \alpha_{\overline{MS}}(1/a)$,
are found to be 3-6 MeV, which have already been included
in the error estimation.

\section{CP-violation in $K$ meson decay}
\label{sec:kcp}
\subsection{$K_0$-$\bar K_0$ mixing parameter $B_K$}
The indirect CP-violation in $K$ meson decays is parameterized by 
$\varepsilon$, which is related to the CKM matrix element as
\beq
{\eta} \left[ (1-{\rho}) A + B \right] \hat B_K =
\frac{\vert {\varepsilon} \vert}{C} ,
\eeqn
where $\rho$ and $\eta$ are Wolfenstein parameters, 
\index{Wolfenstein parameters}
and $A$, $B$ and $C$ are numerical constants.
Here $\hat B_K$ is the renormalization group invariant form of $B_K$,
the $K_0$-$\bar K_0$ mixing parameter,
which is defined as the $K$ meson matrix element of the
$\Delta S = 2$ four quark operator:
\beq
B_K =\frac{\langle \bar K_0 \vert \bar s \gamma_\mu (1-\gamma_5)d\cdot
\bar s \gamma_\mu (1-\gamma_5)d \vert K_0 \rangle}
{\displaystyle 
\frac{8}{3} \langle \bar K_0 \vert \bar s \gamma_\mu (1-\gamma_5)d \vert
0\rangle \langle 0 \vert \bar s \gamma_\mu (1-\gamma_5)d \vert
K_0\rangle }.
\eeqn
Thus the lattice calculation of $B_K$ provides an 
important constraint on the CKM matrix.

\begin{figure}[tbh]
\begin{center}
\epsfig{file=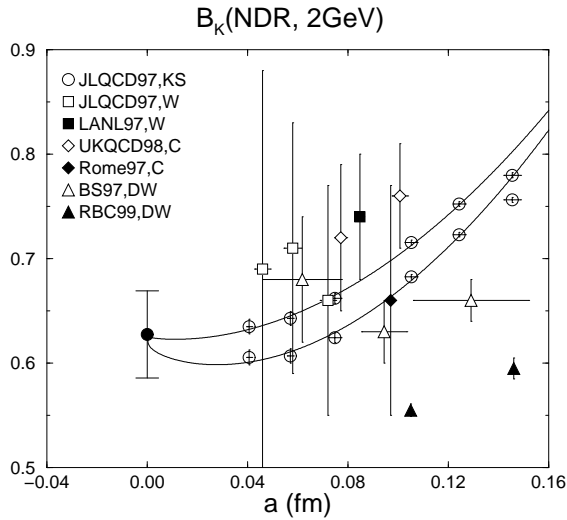, height = 7.0cm}
\caption{$B_K$(NDR, 2 GeV) in quenched QCD
with KS, Wilson/Clover and Domain-wall quark actions,
as a function of the lattice spacing $a$.
Two solid lines represent the continuum extrapolation
by simultaneously fitting
$B_K$ from two different operators of the KS quark,
with the constraint that two agree in the continuum limit.}
\label{fig:bk}
\end{center}
\end{figure}

In Fig.~\ref{fig:bk} we summarize the quenched lattice results for $B_K$(NDR,
2 GeV), renormalized at 2 GeV in the $\overline{MS}$ scheme with 
the naive dimensional regularization (NDR), as a function of lattice 
spacing. 
These results have been obtained with a variety of lattice quark actions. 
The Wilson quark action\cite{wilson,jlqcd_wbk,lanl_wbk} 
\index{Wilson quark action}
breaks chiral symmetry explicitly, while 
the improved Wilson quark action 
(Clover or SW quark action)\cite{sw,ukqcd_cbk,rome_cbk} 
\index{Clover quark action}
reduces the magnitude of the chiral symmetry breaking. 
On the other hand, the Kogut-Susskind (KS) quark action\cite{ks} 
\index{KS quark action}
retains $U(1)$ subgroup of chiral symmetry, which is sufficient to 
guarantee the correct chiral behavior of the $B_K$ parameter.  
Finally the domain-wall(DW) quark action\cite{dwf}, \index{DW quark action}
recently developed and employed in lattice QCD 
calculations\cite{bs_dwbk,rbc_dwbk},
maintains both chiral and flavor symmetries 
at the expense of introducing the extra dimension.
After taking the continuum limit all formulation should give the same answer.

As seen from the figure, 
the result from the KS fermion approach of the JLQCD 
Collaboration\cite{jlqcd_ksbk}
is the most extensive and statistically accurate. 
They employ two different lattice operators for $B_K$, and matching to 
the continuum operator is made with the one-loop renormalization factor.
In order to take into account the 2-loop ambiguity, in addition to the $a^2$ 
scaling violation of the KS quark action, they fit the 
results of $B_K$ from two operators simultaneously by the form
\beq
B_K(a) = B_K + b\ a^2 + c\ \alpha_{\overline{MS}}(1/a)^2 ,
\label{eq:bkfit}
\eeqn
with a unique value of $B_K$ in the continuum limit.
In the figure the fit is represented by solid lines.
By this procedure the JLQCD collaboration obtain
\beq
B_K({\rm NDR, 2 GeV} )=0.628(42) ,
\eeqn
which corresponds to $ \hat B_K = 0.87(6)$.
The error in the continuum limit is much larger than those 
of individual data at non-zero $a$, due to the two-loop ambiguity $\alpha^2$
in eq.~(\ref{eq:bkfit}).  More accurate results requires 
precision determination of the renormalization factor in some
non-perturbative way. 

So far the degenerate quark masses,
$m_s = m_d$, have to be used to avoid a quenched pathology that
$B_K$ diverges in the limit that $m_s \not= m_d\rightarrow 0$.
An 4 to 8\% increase of $B_K$ is predicted by chiral perturbation theory
\index{chiral perturbation theory} ($\chi$PT) for non-degenerate, 
physical quark masses\cite{sharpe}.
As far as the quenching error is concerned,
the value of $B_K$ in 3 flavor full QCD is found to increase by 5 \%
\cite{KPV}. However this should be considered as a preliminary estimate,
since no chiral and continuum extrapolation have been made
in the calculation, and the degenerate quark masses are still used. 

\subsection{Direct CP-Violation}
The direct CP-violation in K meson decays parameterized by 
$\varepsilon^\prime$ is important since a non-zero value of
$\varepsilon^\prime$ strongly supports that the complex phase of the CKM matrix
is the source of CP-violation in the Standard Model.
Recent experiments consistently suggest that $\varepsilon^\prime/\varepsilon$
is non-zero, and the world average of 
experimental values is now 21.3(2.8)$\times 10^{-4}$\cite{KTeV}.

In order to see whether this value is consistent with the standard model
prediction or suggests new physics,
one has to theoretically calculate $\varepsilon^\prime/\varepsilon$ 
within the framework of the Standard Model.
An approximate formula takes the form\cite{buras99}
\beqa
\varepsilon^\prime/\varepsilon & \simeq & {\rm Im} (\lambda_t)  
\times
13\cdot\left[\frac{110\mbox{MeV}}{m_s(2\mbox{GeV})}\right]^2 \cr
&\times&
\left[B_6^{1/2}(1-\Omega_{\eta+\eta^\prime})-0.4\cdot B_8^{3/2}
\left(\frac{m_t}{165\mbox{GeV}}\right)^{2.5}\right]
\left(\frac{\Lambda_{\overline{MS}}^{(4)}}{340\mbox{MeV}}\right),
\eeqan
where $B_{6(8)}^{1/2(3/2)}$ is the matrix element of the
QCD (electroweak) penguin operator in the $\Delta I = 1/2 (3/2)$ $K$ meson
decay, normalized by the vacuum insertion estimate, 
$\lambda_t = V_{td}V_{ts}^*$, and
$\Omega_{\eta + \eta'}$ is the suppression factor by the isospin 
breaking effect
in the quark masses. 
Since this formula indicates that the 
$\varepsilon^\prime/\varepsilon$ increases if the strange quark mass decreases,
it is interesting to calculate its value using the lattice result of $m_s$ 
obtained in full QCD, which is smaller than the previous estimate.

We have evaluated $\varepsilon^\prime/\varepsilon$
employing a more precise formula\cite{buras99} and
taking the standard estimate of parameters that
$ B_6^{1/2} = 1.0(3)$, $B_8^{3/2}=0.8(2)$,
$\Lambda_{\overline{MS}}^{(4)}=340(50)$ MeV, Im$\lambda_t = 
1.33(14)\times 10^{-4}$ and $\Omega_{\eta + \eta'} = 0.25(8)$.
For comparison we use two values of the strange quark mass, 
$m_s^{\overline{MS}} (m_c)$ = 100 (20) MeV from the lattice result
and 130(25) MeV often employed in phenomenology.
Here we conservatively add 5 MeV to the error of lattice result.
 
We list the result in Table~\ref{tab:CP} varying the strange quark mass 
within the estimated error band; the other parameters are
set to be their central values.  
Taking ambiguities in other parameters into account,
the estimation in the Standard Model for $\varepsilon^\prime/\varepsilon$,
for a small strange quark mass indicated by the recent 2 flavor lattice QCD
calculations, is still consistent with the experimental value.

\begin{table}[htb]
\begin{center}
\begin{tabular}{|l|l||l|l|}  
\hline
\multicolumn{2}{|c||}{$m_s(m_c)=100(20)$ MeV: this talk}
& \multicolumn{2}{c|}{$m_s(m_c)=130(25)$ MeV:Ref.~\protect{\cite{buras99}} }\\
\hline
$m_s(m_c)$ & $\varepsilon^\prime/\varepsilon$ &
$m_s(m_c)$ & $\varepsilon^\prime/\varepsilon$ \\
\hline
80 MeV & $21.2\times 10^{-4}$ & 105 MeV & $11.5\times 10^{-4}$ \\
100 MeV & $13.0\times 10^{-4}$ & 130 MeV & $7.0\times 10^{-4}$ \\
120 MeV & $8.5\times 10^{-4}$ & 155 MeV & $4.4\times 10^{-4}$ \\
\hline
\end{tabular}
\caption{The strange quark mass and $\varepsilon^\prime/\varepsilon$}
\label{tab:CP}
\end{center}
\end{table}

It should be noted
that only the strange quark mass is varied here while other parameters
are kept fixed.  This has inherent uncertainties since 
other parameters such as $B_6^{1/2}$ and $B_8^{3/2}$ also depend on the 
strange quark mass.
(See ref.~\cite{DGGM} for a quark mass independent parameterization of 
these matrix elements.)
For a complete prediction for $\varepsilon^\prime/\varepsilon$ from the
Standard Model, we need to evaluate $B_6^{1/2}$ and $B_8^{3/2}$ 
also by lattice calculations. This possibility will be considered next.

The direct calculation of the $K\rightarrow\pi \pi$ amplitude, in particular
for the $\Delta I =1/2$ case, is notoriously difficult,
mainly due to the euclidean nature of space-time on the lattice\cite{MT}.
Instead of the direct calculation, one may evaluate 
$K\rightarrow \pi$ amplitudes, which are related
to the $K\rightarrow\pi \pi$ amplitude at the lowest order of
$\chi$PT as follows\cite{bdspw}.
\beq
\langle \pi^+\pi^-\vert O_i \vert K^0\rangle =
\langle \pi^+ \vert O_i- \alpha_i O_{sub} \vert K^+ \rangle
\cdot \frac{m_K^2-m_\pi^2}{m_M^2 f_M} ,
\eeq{eq:cpt}
where $O_i$ is a $\Delta S =1$ four quark operator,
and $\alpha_i$ is determined by
\begin{equation}
0 = \langle 0\vert O_i- \alpha_i O_{sub} \vert K^0 \rangle
\end{equation}
with  $ O_{sub} \equiv (m_d+m_s) \bar s d + (m_d-m_s) \bar s\gamma_5 d$,
which can mix with $O_i$ under operator renormalizations.
Here $m_K$ and $m_\pi$ are physical $K$ and $\pi$
meson masses, while $m_M$ is an unphysical degenerate $K$ and $\pi$ meson 
mass, used in the lattice calculation of the $K\rightarrow\pi$ matrix element.
Therefore the left-hand side of eq.~(\ref{eq:cpt}) should be
independent on $m_M$ at the lowest order of $\chi$PT. 
For this method to work, chiral symmetry of the lattice quark action 
is crucially important.

\begin{figure}[htb]
\begin{center}
\epsfig{file=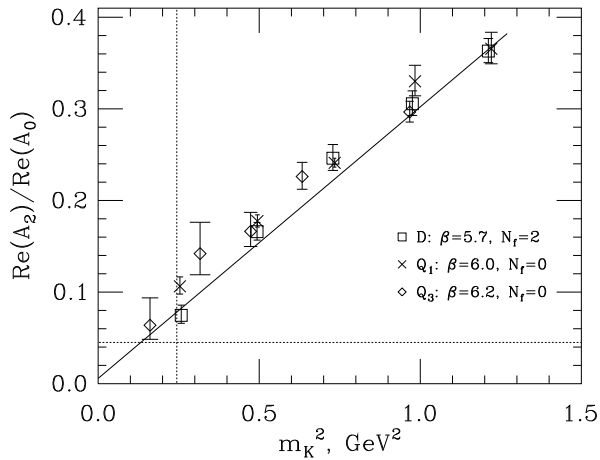, height = 6.0cm}
\caption{Re($A_2$)/Re($A_0$) in quenched QCD with KS quark action
at $\beta =6.0$(crosses) and $\beta = 6.2$(diamonds),
as a function of the (unphysical) $K$ meson mass squared, together with the one
in full QCD at $\beta=5.7$( suqares). A horizontal dotted line represents
the experimental value of the ratio, while the vertical one is the
experimental value of $K$ meson mass squared.}
\label{fig:a2a0ks}
\end{center}
\end{figure}

Recently, applying this method with the KS quark action\cite{kp_kpp} 
or the domain-wall quark action\cite{rbc_dwbk,rbc_kpp}, 
both of which have good chiral properties,
statistically meaningful results have been obtained
for $K \rightarrow\pi\pi$ amplitudes, relevant for the $\Delta = 1/2$
rule. In Fig.~\ref{fig:a2a0ks}, Re ($A_2$)/Re ($A_0$), where
$A_I$ is the $K\rightarrow\pi\pi$ amplitude for the final $\pi$-$\pi$ state 
with the isospin $I$, is plotted as a function of $m_M^2$ 
(denoted as $m_K^2$ in the figure),
in both quenched and full QCD with the KS quark action\cite{kp_kpp}.
Although the ratio is still larger than the experimental value 
($\Delta I =1/2$ enhancement is still smaller than the experimental one)
and the strong $m_M^2$ dependence, mainly caused by Re $(A_2)$,
is observed, 
the result suggests that the method is useful. 
Calculations with better control
over systematic errors as well as statistical errors
will be the next step. A similar result for Re $A_0$ 
with small $m_M^2$ dependence
is recently reported, by employing domain-wall quarks in the quenched 
approximation\cite{rbc_dwbk,rbc_kpp}.

Although results for $\varepsilon'/\varepsilon$ have also been 
obtained by the same  method with the KS quark action, 
it has been found that effects associated with
1-loop renormalization of weak operators are too large to obtain a reliable
estimate of $\varepsilon'/\varepsilon$\cite{kp_kpp}.
Partially employing non-perturbative renormalization factors, a negative 
value of $\varepsilon'/\varepsilon$ is 
obtained\cite{kp_kpp}. 
Recently a large negative value of $\varepsilon'/\varepsilon$ is also reported
from the calculation by domain-wall quarks with 
non-perturbative renormalization factors\cite{rbc_dwbk,rbc_kpp}.
The value of $B_6^{1/2}$ is also negatively large.
However, further confirmations of these results, 
which include a systematic study of scaling violations for the continuum
extrapolation and the validity of the reduction method via $\chi$PT,
will be needed for establishing a lattice estimate of 
$\varepsilon'/\varepsilon$.

For the $\Delta I =3/2$ decay, on the other hand, 
the estimation by the lowest order of the $\chi$PT gives 
reasonable result. For example,
the quenched results of $B_7^{3/2}$ and $B_8^{3/2}$ 
are given in Table~\ref{tab:Bs}.
Furthermore the direct calculation of the $\Delta I =3/2$ $K\rightarrow\pi\pi$
decay amplitude successfully gives
$\langle \pi^+\pi^0\vert Q_+\vert K^+\rangle =$ (8.9--10.6) $\times 10^{-3}$
GeV$^3$ in the quenched approximation\cite{jlqcd_a2}, in good agreement with
the experimental value 10.4$\times 10^{-3}$ GeV$^3$.
\begin{table}[tbh]
\begin{center}
\begin{tabular}{|l|l|l|l|l|l|}  
\hline
References & action & $a$ & Renormalization & $B_7^{3/2}$(2 GeV) &
$B_8^{3/2}$(2 GeV) \\
\hline
GBS96\protect{\cite{lanl_wbk}} & Wilson & $\not= 0$ & Perturbative &
0.58(2)$(^7_3)$ & 0.81(3)$(^3_2)$ \\
KGS97\protect{\cite{kgs97}} & KS & $\rightarrow 0$ & Perturbative &
0.62(3)(6) & 0.77(4)(4) \\
Rome97\protect{\cite{rome_cbk}} & Clover & $\not= 0$ & Non-Perturbative &
0.72(5) & 1.03(3) \\
Rome97\protect{\cite{rome_cbk}} & Clover & $\not= 0$ & Perturbative &
0.58(2) & 0.83(2) \\
UKQCD98\protect{\cite{ukqcd_cbk}} & Clover & $\not= 0$ & Perturbative &
0.58$(^5_4)(^2_8)$ & 0.80(8)$(^1_4)$ \\
\hline
\end{tabular}
\caption{Lattice estimates for $B_7^{3/2}$ and $B_8^{3/2}$
in the quenched approximation.}
\label{tab:Bs}
\end{center}
\end{table}

\section{Weak matrix elements of heavy-light mesons}
\label{sec:heavy}

\subsection{Lattice action for heavy quarks}

A practical problem for lattice study of heavy quark physics
is that the cut-off $1/a$ in the current simulations, which is typically
$1\sim 4$ GeV, is smaller than the $b$ quark mass, $m_b \simeq 4$ GeV.
Therefore the discretization error of order $m_ba$ is 
expected to be large. 
If the Wilson/Clover action is used for the heavy quark,
the numerical simulations has to be restricted to the charm quark mass region,
to avoid the large discretization errors, and the results 
has to be extrapolated to the $b$ quark mass,
with the help of results in the static limit.
We call this method as the extrapolation method below.

It has been pointed out\cite{fnal}, however, that direct simulations 
at the $b$ quark mass is possible if
the large discretization errors is removed by a careful interpretation of 
the results.
We call this the direct method.

Although the continuum limit can be taken in both methods,
the $a$ dependence of the results is complicated and non-linear in the latter.

If the non-relativistic formulation of quark is used instead,
one can directly simulate the $b$ quark more easily and accurately.
This method, called lattice NRQCD\cite{nrqcd}, \index{NRQCD}
is an effective theory defined by an inverse expansion in the heavy quark mass.
Taking the continuum limit is not possible with this method, and 
the systematic error increases for the charm quark.

Since there is no best method so far,
one always has to check that different methods
gives consistent results within estimated errors.

\subsection{Leptonic decay constants}
\begin{figure}[htb]
\begin{center}
\epsfig{file=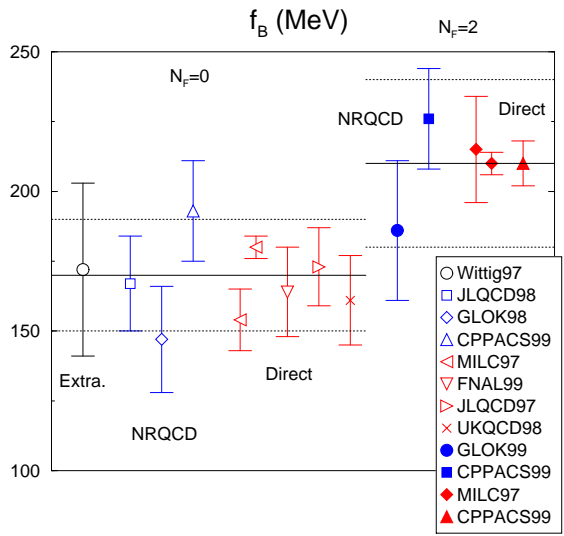, height = 7.0cm}
\caption{$f_B$, from different methods 
in quenched (open) and full (filled) QCD.}
\label{fig:fb}
\end{center}
\end{figure}
One of the simplest quantities in weak matrix elements
is the leptonic decay constant of a heavy-light meson,
which is important to determine $V_{td}$
and $V_{ts}$ from neutral $B$ meson mass differences:
\beq
\Delta M_q = \frac{G_F M_W^2}{6\pi^2}\eta_{B_q} S( m_t/M_W)
f_{B_q}^2 \hat B_{B_q} \vert V_{tq}\vert^2 ,
\label{eq:dmb}
\eeqn
where $f_{B_q}$ denotes the decay constant of the $B_q$ meson.
In particular 
a significant effort among lattice groups
has been devoted to the calculation of
$f_B \equiv f_{B_d}$, which has not been experimentally measured yet.

After many years of calculations,
results for $f_B$ from several groups with different methods
are gradually converging. In Fig.~\ref{fig:fb}, the summary of the
latest results for $f_B$ form different methods, in both quenched ($N_f=0$)
and full ($N_f=2$) QCD, are summarized. 

Results in quenched QCD, represented by open symbols, from
the extrapolation method\cite{wittig}, 
NRQCD\cite{jlqcd_nrfb,glok_nrfb,cppacs_nrfb} and the direct 
method\cite{milc_fnfb,fnal_fnfb,jlqcd_fnfb},
are consistent within 10 \% errors, showing that the lattice
methods currently employed indeed work for $b$ quark.
Recent calculations in full QCD with 2 flavors of dynamical 
quarks ($N_f=2$),
denoted by filled symbols,
from NRQCD\cite{glok_nrfb_full,cppacs_nrfb} and
the direct method\cite{milc_fnfb,milc_fnfb_full,cppacs_fnfb_full},
indicate that $f_B$ increases by 10--20 \%.

\begin{table}[htb]
\begin{center}
\begin{tabular}{|l|l|l|l|}  
\hline
Quantity & $N_f=0$ & $N_f=2$ & Experiments \\
\hline
$f_B$ & 170(20) MeV & 210(30) MeV & \\
$f_{B_s}$ & 195(20) MeV & 245(30) MeV & \\
$f_D$ & 200(20) MeV &  & $\le$ 310 MeV\\
$f_{D_s}$ & 220(20) MeV & & 241(21)(30) MeV \\
\hline
$f_{B_s}/f_B$ & 1.15(4) & 1.16(4) & \\
$f_{D_s}/f_D$ & 1.10(6) &         & \\
\hline
\end{tabular}
\caption{Lattice estimates for leptonic decay constants in
$N_f=0$ and 2.}
\label{tab:fb}
\end{center}
\end{table}

A summary of leptonic decay constants is given in Table~\ref{tab:fb}.
The present best estimate of $f_B$ in 2 flavor QCD is 210(30) MeV, which is
larger than the previous quenched result.
The ratio, $f_{B_s}/f_B$, which has less systematic uncertainty,
on the other hand, remains unchanged from quenched QCD to full QCD.
This quantity will become important if both $\Delta M_d$ and $\Delta M_s$ are
experimentally measured.

\subsection{ $B$ meson mixing parameter $B_B$}

As is seen in eq.~(\ref{eq:dmb}),
the mixing parameters of neutral $B_q$ mesons $B_{B_q}$ ($q=d,s$),
defined by
\beq
B_{B_q}(\mu)=\frac{\langle \bar B_q \vert \bar b \gamma_\mu (1-\gamma_5) q\cdot
\bar b \gamma_\mu (1-\gamma_5) q \vert B_q \rangle}
{\displaystyle\frac{8}{3} \langle \bar B_q \vert \bar b 
\gamma_\mu (1-\gamma_5)q \vert
0\rangle \langle 0 \vert \bar b \gamma_\mu (1-\gamma_5)q \vert
B_{B_q}\rangle } ,
\eeqn
are important to determine $V_{td}$ and $V_{ts}$ 
from $\Delta M_q$, together with decay constants $f_{B_q}$.

\begin{figure}[tb]
\begin{center}
\epsfig{file=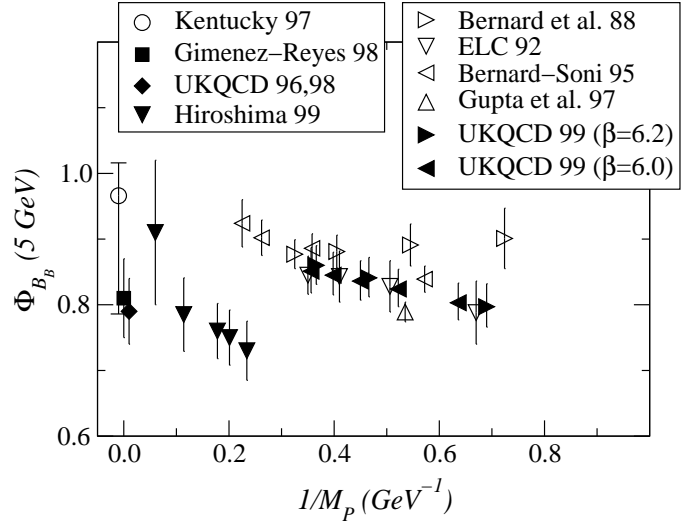, height = 6.0cm}
\caption{A compilation of quenched results for
$\Phi_{B_B}$(5 GeV) as a function of
a heavy meson mass inverse, $1/M_P$.}
\label{fig:BB}
\end{center}
\end{figure}

Lattice results so far are obtained within the quenched approximation 
at fixed lattice spacings. 
A compilation of lattice results for
$\Phi_{B_B}(\mu )\equiv (\alpha_s (M_p)/\alpha_s(M_B))^{2/\beta_0}
B_B(\mu )$ at $\mu =$ 5 GeV is shown in Fig.~\ref{fig:BB},
as a function of a heavy meson mass inverse, $1/M_P$.
The coupling factor is introduced to cancel the $\ln(M_P)$
dependence in the $B$ parameter\cite{LL}.

Data in the figure are classified into 3 distinct groups.
The first is the calculation from the static-light approximation,
where the heavy meson mass is fixed to be infinite. The results are
consistent among the Kentucky group\cite{kent_bb},
Gimenez-Reyes\cite{gm_bb,gr_bb} and UKQCD\cite{ukqcd_bb},
though the error of the Kentucky group 
is rather large due to the ambiguity of 1-loop renormalization in 
Wilson quark action.
Therefore, we take an average over results of the latter two 
groups, who employ the Clover quark action,  as the current lattice estimate,
giving
$B_{B_d}(m_b) = 0.80(5)$ ($\hat B_{B_d}$ = 1.28(8)).

The second result comes from a NRQCD-light method by the Hiroshima 
group\cite{hiroshima_bb}.
They observe a heavy meson mass dependence: $B_B$ decreases as $B$ meson mass
decreases from the static limit. 
As is seen from the figure, their result in the static limit
is reasonably consistent with the previous result from static-light
calculations.

There are several calculations using the extrapolation method. They are
rather old results from Wilson quark 
action\cite{bernard_bb,elc_bb,bs_bb,lanl_wbk}, except UKQCD's this year
with the Clover action\cite{ukqcd_clbb}. Although the results are
internally consistent, they seem to differ from the static result
in the static limit and the NRQCD result at the $B$ meson mass,
as is seen in the figure.
More work is needed to reduce this uncertainty and
to obtain the reliable estimate for $B_B$.

We take the static result as our tentative estimate for $B_B$,
since systematic errors are best controlled in this method so far:
\beq
B_{B_d}(m_b) = 0.80(5), \qquad ( \hat B_{B_d}=1.28(8) ), \qquad
\xi_{sd}
\equiv \displaystyle \frac{f_{B_s} \sqrt{B_{B_s}}}{f_{B_d} \sqrt{B_{B_d}}}
= 1.17(6) .
\eeqn
Combining this value with the previous estimate of $f_B$, one obtains
\beq
\sqrt{\hat B_{B_d}} f_{B_d} = 240 (36) {\rm MeV} .
\eeqn

\subsection{Form factors of semi-leptonic decays}
In this section form factors of semi-leptonic decays of $B$ or $D$ mesons
are considered. 
For pseudo-scalar to pseudo-scalar decays there are two form factors,
$f^+(q^2)$ and $f^0(q^2)$, which are defined as
\beq
\langle P(k)\vert V^\mu \vert H(p) \rangle =
{f^+(q^2)}\left[ (p+k)^\mu-\frac{m_H^2-m_P^2}{q^2}q^\mu\right]
+{f^0(q^2)}\frac{m_H^2-m_P^2}{q^2}q^\mu ,
\eeq
where $q=p-k$ is the momentum transfer, $H = B$ or $D$, and
$P= D$, $K$, $\eta$ or $\pi$.
Four form factors for pseudo-scalar to vector decays,
$V(q^2)$, $A_{1,2}(q^2)$ and $A(q^2)$ are defined as
\begin{eqnarray*}
\langle V(k,\varepsilon) \vert V^\mu \vert H(p)\rangle &=&
\frac{ 2 { V(q^2)}}{m_H+m_V} \epsilon^{\mu\nu\alpha\beta}p_\nu k_\alpha
\varepsilon^*_\beta \\
\langle V(k,\varepsilon)\vert A^\mu \vert H(p) \rangle & =&
i(m_H+m_V){ A_1(q^2)}\ \varepsilon^{*\mu}
-i\frac{{ A_2(q^2)}}{m_H+m_V}\varepsilon^*\cdot p\ (p+k)^\mu \\
& &+i\frac{{ A(q^2)}}{q^2} 2 m_V \varepsilon^*\cdot p\ q^\mu 
\end{eqnarray*}
where $V = D^*$, $K^*$, $\phi$ or $\rho$.

\subsubsection{$D\rightarrow K^{(*)}\ell\nu$, $\pi(\rho) \ell \nu$}
We first consider $D$ meson decays to light mesons, 
$K$, $K^*$, $\pi$ or $\rho$.
The purpose of this calculation is two-fold:
Assuming that $V_{cs}= 1-\lambda^2/2$, one can establish the validity
of lattice calculations for semi-leptonic form factors.
Conversely one may extract $V_{cs}$ or $V_{cb}$ directly from experimental
data, using the lattice estimate of the form factors.

The calculation of these form factors is easier than others,
because direct simulations at $m_c$ or $m_s$ can avoid subtlety of the
quark mass extrapolation. In addition the fact that all physical 
as well as unphysical 
$q^2$ regions are covered makes interpolation to $q^2=0$ possible and easy.

\begin{figure}[htb]
\begin{center}
\epsfig{file=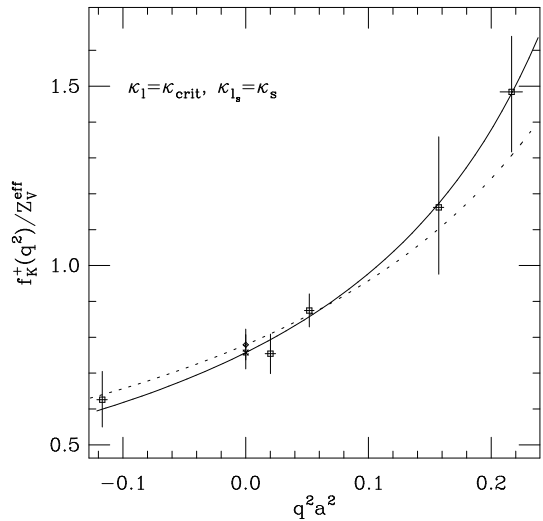, height = 6.5cm}
\caption{The form factor $f^+(q^2)$ for $D\rightarrow K\ell\nu$ decay,
as a function of $q^2$.}
\label{fig:DtoK}
\end{center}
\end{figure}

As an example, in Fig.~\ref{fig:DtoK}
a lattice result\cite{ukqcd_d2k}, already in 1995, for 
the form factor $f^+(q^2)$ of the $D \rightarrow K\ell\nu$ decay, 
is given as a function of $q^2$ in quenched QCD.
As this figure shows, lattice methods work well for these decays.
For a summary of quenched estimates for these form factors at $q^2 =0$,
see Ref.\cite{FS}.

\begin{figure}[bht]
\begin{center}
\epsfig{file=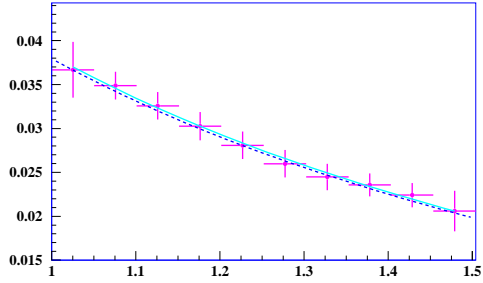, height = 5.0cm}
\caption{${\vert \cal F}_{B\rightarrow D^*}(\omega) V_{cb}\vert$
as a function of $\omega$ from DELPHI collaboration.}
\label{fig:delphi}
\end{center}
\end{figure}
\subsubsection{$B\rightarrow D^{(*)}\ell\nu$}
The large branching ratio of $B\rightarrow D^{(*)}\ell\nu$ decays
makes a determination of $\vert V_{cb} \vert$ very precise.
Differential decay rates are related to
$\vert V_{cb} \vert^2$ and the form factor ${\cal F}_{B\rightarrow D}$ 
by
\beq
\frac{d\Gamma}{d\omega}(B\rightarrow D^{(*)}\ell\nu)
=({\rm known \  factor})\times
\vert V_{cb}\vert^2 \vert{\cal F}_{B\rightarrow D^{(*)} }
(\omega)\vert 
\eeqn
where $\omega = v_B\cdot v_{D^{(*)}}$ is the velocity transfer.
For example, the very accurate result for
$\vert {\cal F}_{B\rightarrow D^{*}} (\omega ) V_{cb}\vert$
from the DELPHI collaboration is shown in Fig.~\ref{fig:delphi}.
Taking the $\omega \rightarrow 1$ limit, 
one obtain $\vert {\cal F}_{B\rightarrow D^{(*)}}
(1) V_{cb}\vert = (37.95\pm 1.34\pm 1.59) \times 10^{-3}$\cite{delphi}.

Now the most important task of lattice QCD is the determination of
$\vert {\cal F}_{B\rightarrow D^{(*)}}(1 )\vert$, 
in order to extract $\vert V_{cb}\vert$ from experimental results.
They are given by
\beqa
{\cal F}_{B\rightarrow D}(1) &=& h_+(1) -\frac{m_B-m_D}{m_B+m_D}
h_-(1) \\
{\cal F}_{B\rightarrow D^*}(1) &=& h_{A_1}(1) ,
\eeqan
in terms of new form factors, defined by
\beqa
\frac{\langle D(v')\vert \bar c \gamma_\mu b \vert B(v)\rangle }{\sqrt{m_Bm_D}}
&=& \left[ {h_+ (\omega)}(v+v')_\mu + {h_-(\omega)}(v-v')_\mu\right]
\\
\frac{\langle D^*(v')\vert \bar c \gamma_\mu\gamma_5 b \vert B(v)\rangle }
{\sqrt{m_Bm_D^*}}
&=&  (\omega +1) \varepsilon^*_\mu {h_{A_1}(\omega)} .
\eeqan
Since the heavy quark symmetry \index{heavy quark symmetry} implies
$
{\cal F}_{B\rightarrow D}(1) = {\cal F}_{B\rightarrow D^*}(1) = 1
$
in the $m_b$, $ m_c \rightarrow \infty$ limit,
the heavy quark mass dependence has to be determined precisely.

\begin{table}[htb]
\begin{center}
\begin{tabular}{|l|l|l|}  
\hline
& Lattice & QCD sum rule \\
\hline
${\cal F}_{B\rightarrow D}(1)$  & { 1.058 (16)$(^{+14}_{-6})$} &
 0.98(7)\\
${\cal F}_{B\rightarrow D^*}(1)$  & { 0.935 (22)$(^{+23}_{-24})$} &
0.91(6)/0.92(4) \\
\hline
\end{tabular}
\caption{Lattice results for 
$\vert {\cal F}_{B\rightarrow D^*}(1)\vert$, together with those from
QCD sum rule.}
\label{tab:f1}
\end{center}
\end{table}

These form factors have been evaluated recently 
by the Fermilab group\cite{fnal_b2d},
using the direct method with Clover quark action,
in quenched approximation at $a^{-1} \sim $ 1 GeV. 
In order to cancel systematic as well as statistical errors of 
lattice matrix elements,
they employ the ratio method, through which
$\vert h_+ (1) \vert^2$, for example, is obtained as
\beq
\vert h_+^{B\rightarrow D}(1)\vert^2
=\frac{ \langle D \vert V_0^{cb}\vert B\rangle
\langle B \vert V_0^{bc}\vert D\rangle}
{\langle D \vert V_0^{cc}\vert D\rangle
\langle B \vert V_0^{bb}\vert B\rangle} .
\eeqn
The precison of the result makes it possible to
determine the heavy quark mass dependence of $h$'s up
to $1/m_{b,c}^3$ for $h_+(1)$ and $1/m_{b,c}^2$ for $h_{-,A_1}(1)$.

The summary of their results, together with results from the QCD sum 
rule approach\cite{SR_b2d},
is given in Table~\ref{tab:f1}.
Already errors of lattice results are comparable to or even smaller than
those of QCD sum rule results.
Since the deviation from 1, not the value itself, is important
for $F_{B\rightarrow D^{(*)}}(1)$, the systematic as well as
the statistical errors should be reduced further.
Improved calculations with smaller lattice spacings
for the continuum extrapolation and/or the inclusion of the full QCD effect
should be performed as the next step.
Results from such calculations, together with experimental data,
will reduce errors of $\vert V_{cb}\vert$ significantly.

\bigskip

As far as the shape of $F_{B\rightarrow D^{(*)}}(\omega)$ is concerned,
the lattice calculation has a longer history than that for  
$F_{B\rightarrow D^{(*)}}(1)$.
It is important to calculate the shape of form factors, 
as a check of theoretical method and for a reduction of errors
in the extrapolation of experimental data.

\begin{figure}[bth]
\begin{center}
\epsfig{file=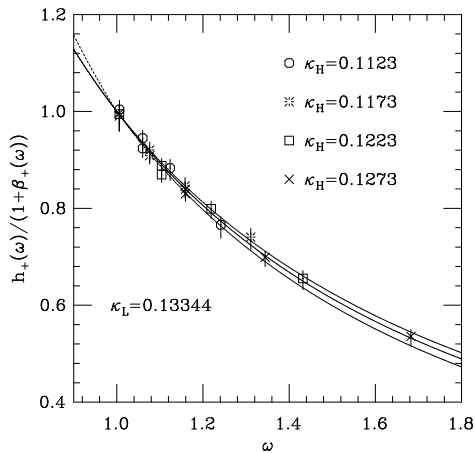, height = 6.5cm}
\caption{The radiatively corrected form factor $h_+(\omega)$
in quenched QCD from UKQCD collaboration. 
Different symbols correspond to different heavy quark masses,
while the light quark mass is kept fixed.}
\label{fig:IW}
\end{center}
\end{figure}

Heavy quark symmetry implies that these form factors are related to one 
universal, mass-independent function, $\xi(\omega)$, the famous Isgur-Wise 
function:
\beqa
h_{+,A_1}(\omega) & =& (1 + \gamma_{+,A_1}(\omega)+\beta_{+,A_1}(\omega))
{\xi(\omega)} ,
\\
h_{-}(\omega) & =& (\gamma_{-}(\omega)+\beta_{-}(\omega)) { \xi(\omega)},
\eeqan
where $\beta_{\pm,A_1}$ represent radiative corrections, while
$\gamma_{\pm,A_1}(\omega)$ represent heavy quark mass dependences.
The Isgur-Wise function $\xi(\omega)$ has already been calculated on the
lattice. For example, the recent result for
$h_+(\omega)/(1+\beta_+(\omega))$ from the UKQCD 
collaboration\cite{ukqcd_b2d} with the Clover quark action is 
given in Fig.~\ref{fig:IW}.
They employ four different heavy quark masses, and find the
mass-independence of $h_+(\omega)/(1+\beta_+(\omega))$, 
which suggests that the correction to the heavy quark mass limit is indeed 
small. Therefore, within errors,  $h_+(\omega)/(1+\beta_+(\omega))$ can be
identified with the Isgur-Wise function, $\xi (\omega )$.
In future investigations  it will be important to extract mass dependent term,
$\gamma_{\pm,A_1}(\omega)$, in order to construct the shape of
$F_{B\rightarrow D^{(*)}}(\omega)$ precisely.

\subsubsection{$B\rightarrow \pi\ell\nu$, $\rho\ell\nu$}
The last quantity of our concern is the form factors of 
$B\rightarrow \pi$, $\rho$ decays, which are rare decays, and
can be used to determine $\vert V_{ub}\vert$.

Lattice calculations for these form factors are restricted
to large $q^2$ regions only, since momenta of $B$ and $\pi/\rho$ have to be
smaller than $1/a$, to avoid large discretization errors.
Therefore experimental data for the partial decay rate 
in the large $q^2$ region are
necessary for such lattice calculations being useful.
Indeed statistics in the recent experiment is precise enough
to attempt an initial study for the partial decay rate\cite{cleo_b2r}
and the quality of data is expected to be further improved.

\begin{figure}[htb]
\begin{center}
\epsfig{file=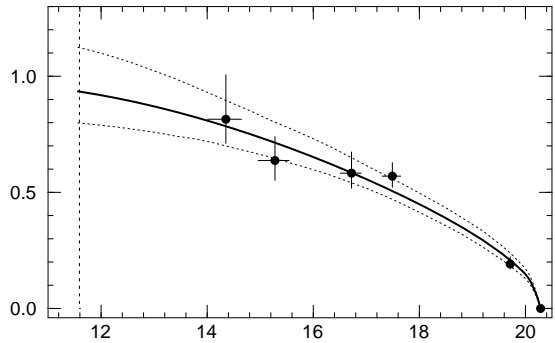, height = 5.0cm}
\caption{The form factor $B\rightarrow \rho\ell\nu$ decay
as a function of $q^2$ in quenched QCD from the UKQCD collaboration.
The solid line represents the fit given in the text, and vertical
dotted line shows $q^2_{\rm max}$ for $B\rightarrow D^{*}$ decay.}
\label{fig:b2rho}
\end{center}
\end{figure}

The UKQCD collaboration calculates the form factor relevant for
the differential decay rate of $B\rightarrow\rho\ell\nu$
in the quenched approximation at $a\not= 0$, using the Clover quark 
action\cite{ukqcd_b2r}.
As shown in Fig.~\ref{fig:b2rho}, 
fitting lattice data near $q^2 = q^2_{\rm max}$ by
the form, 
\beq
\displaystyle \frac{d\Gamma (B\rightarrow\rho\ell\nu)}{dq^2} 
\frac{10^{12}}{\vert V_{ub}\vert^2}
=
{c^2}(1+{ b}(q^2-q_{\rm max}^2))
\times(\mbox{phase space factor}) .
\eeqn
they obtain
\beq
c = 4.6 \pm 0.7 {\rm GeV} \qquad b = ( -8 ^{+4}_{-6})\times 10^{-2}
{\rm GeV}^{-2} .
\eeqn
As the precision of the differential decay rate will be further improved
in future experiments, this information will be useful in the
extraction of $\vert V_{ub}\vert$.

There are two form factors for $B\rightarrow \pi\ell\nu$ decay,
defined as
\beq
\langle \pi (k)\vert \bar u \gamma_\mu b \vert B(p) \rangle =
{f^+(q^2)}\left[ (p+k)_\mu-\frac{m_B^2-m_\pi^2}{q^2}q_\mu\right]
+{f^0(q^2)}\frac{m_B^2-m_\pi^2}{q^2}q_\mu .
\eeqn
Only $f^+(q^2)$ contributes to the differential decay rate as
\beq
\frac{d\Gamma}{d q^2}(B\rightarrow \pi\ell\nu)
\propto {\vert V_{ub}\vert^2} {\vert f^+(q^2)\vert^2},
\eeqn
while the contribution of $f^0(q^2)$ is negligible in the decay rate since
$ q_\mu L_\mu \propto m_l$, where $L_\mu$ is the leptonic weak current and
$m_l$ is the lepton mass.
However $f^0(q^2)$ may be used to check lattice calculations
by the relation that $f^0(q^2_{\rm max})={f_B}/{f_\pi}$,
implied by the soft pion theorem in the $m_\pi\rightarrow 0$ limit.

It has been found that the calculation of these form factors is 
rather difficult.
The chiral extrapolation of $f^0(q^2)$
linear in the light quark mass does not appear to satisfy the relation
of the soft pion theorem\cite{lat99,lat97}.
Although it is claimed that the inclusion of the a square root term
in the chiral extrapolation solves the problem\cite{ukqcd_b2p}, 
at this moment we have to conclude that 
lattice results are premature for a detailed comparison with experimental data,
and that more work will be needed on this form factor.

\section{Impact on determination of CKM matrix}
\label{sec:ckm}
\begin{table}[htb]
\begin{center}
\begin{tabular}{|l|l||l|l|l|}
\hline
\multicolumn{2}{|c||}{Experimental inputs} &
\multicolumn{3}{c|}{Theoretical inputs} \\
\hline
quantities & value & quantities & { Buras99} & { Lattice} \\
\hline
$\vert V_{cb}\vert$ & 0.040(2) & $\hat B_K$ & 0.80(15) & { 0.87(6)} \\
$\vert V_{ub}\vert$ & $3.56(56)\cdot 10^{-3}$ 
& $f_B$ &               & { 210(30)} MeV \\
$\vert \varepsilon\vert$ & $2.280(13)\cdot 10^{-3}$ & 
$\hat B_{B_d}$ &      & { 1.28(0.08)} \\
$\Delta M_d $ & 0.471(16) ps$^{-1}$ & 
$\sqrt{\hat B_B}f_B$ & 200(40) MeV & { 240(23)} MeV \\
$\Delta M_s $ & $>$ 12.4 ps$^{-1}$ & 
$\xi$   & 1.14(8) & { 1.17(6)} \\
$m_t$ & 165(5) GeV &  & & \\
\hline
\end{tabular}
\caption{Input parameters.}
\label{tab:input}
\end{center}
\end{table}

Let us collect the lattice results and 
discuss their impact on the CKM matrix determination.
We employ the four standard constraint relations given below to determine
the unitarity triangle (see ref.~\cite{buras99} for detailed notations).
\beqa
 \sqrt{{\bar\rho^2} +{\bar\eta^2}}
&=& (1-\frac{\lambda^2}{2})\frac{1}{\lambda}
\left\vert \frac{ V_{ub} }{ V_{cb} } \right\vert ,
\label{eq:ckm1} \\
\frac{\vert {\varepsilon} \vert}{\lambda^{10} C_\varepsilon}
&=&
{\bar \eta} \left[ (1-{\bar\rho})A^2\eta_2 S_0(x_t) + P_0(\varepsilon)
\right] A^2 \hat B_K ,
\label{eq:ckm2} \\
\sqrt{(1-{\bar\rho})^2+{\eta^2}} &=& \frac{1}{\lambda}
\left\vert\frac{V_{td}}{V_{cb}}\right\vert , 
\cr
\mbox{ with }
& \Delta M_d &  = \frac{G_F^2}{6\pi^2}\eta_B\cdot m_B\cdot \sqrt{\hat B_{B_d}}
f_{B_d} m_W^2 \ S_0(x_t)\  \vert {V_{td}}\vert^2,
\label{eq:ckm3} \\
\sqrt{(1-{\bar\rho})^2+{\eta^2}} & \ge &
\xi \sqrt{\frac{10.2/ps}{(\Delta M_s)_{min}}} ,
\eeqa{eq:ckm4}
where $\bar\rho = \rho ( 1 - \lambda^2/2)$ and
$\bar\eta =\eta ( 1-\lambda^2/2)$.
Experimental inputs, which come from $b\rightarrow u$ transition
for eq.~(\ref{eq:ckm1}),
$K_0 - \bar K_0$ ($\varepsilon$) for eq.~(\ref{eq:ckm2}), 
and $B_0 - \bar B_0$ mixings for eqs.~(\ref{eq:ckm3}, \ref{eq:ckm4}),
are summarized in Table~\ref{tab:input}. 
As theoretical inputs, we consider two choices, also  specified 
in Table~\ref{tab:input}: one choice based on the lattice results 
summarized in this review, and the other choice from Ref.\cite{buras99}. 
We shall refer to the latter as a standard one. 

\begin{figure}[htb]
\begin{center}
\epsfig{file=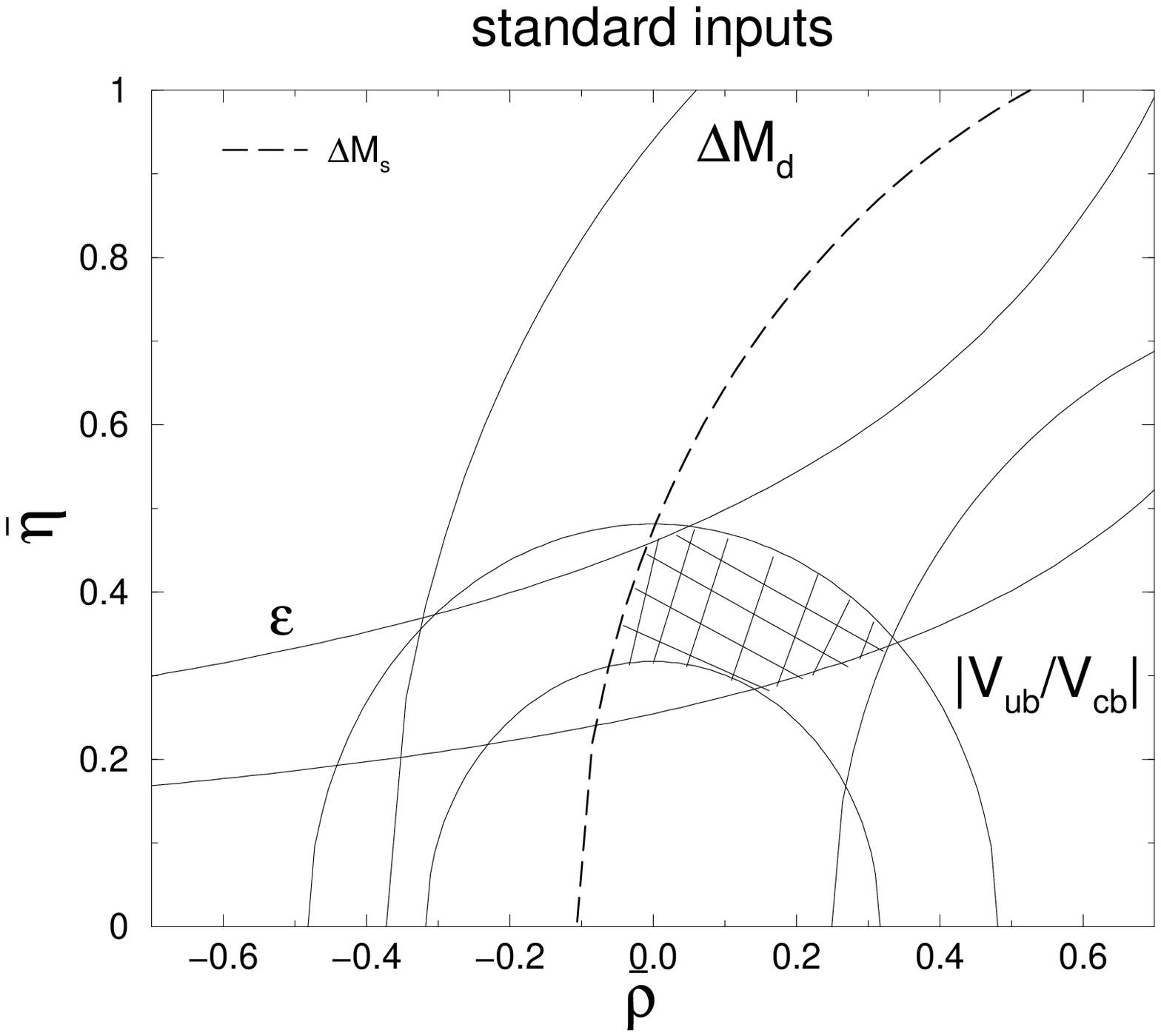, height = 5.7cm}
\epsfig{file=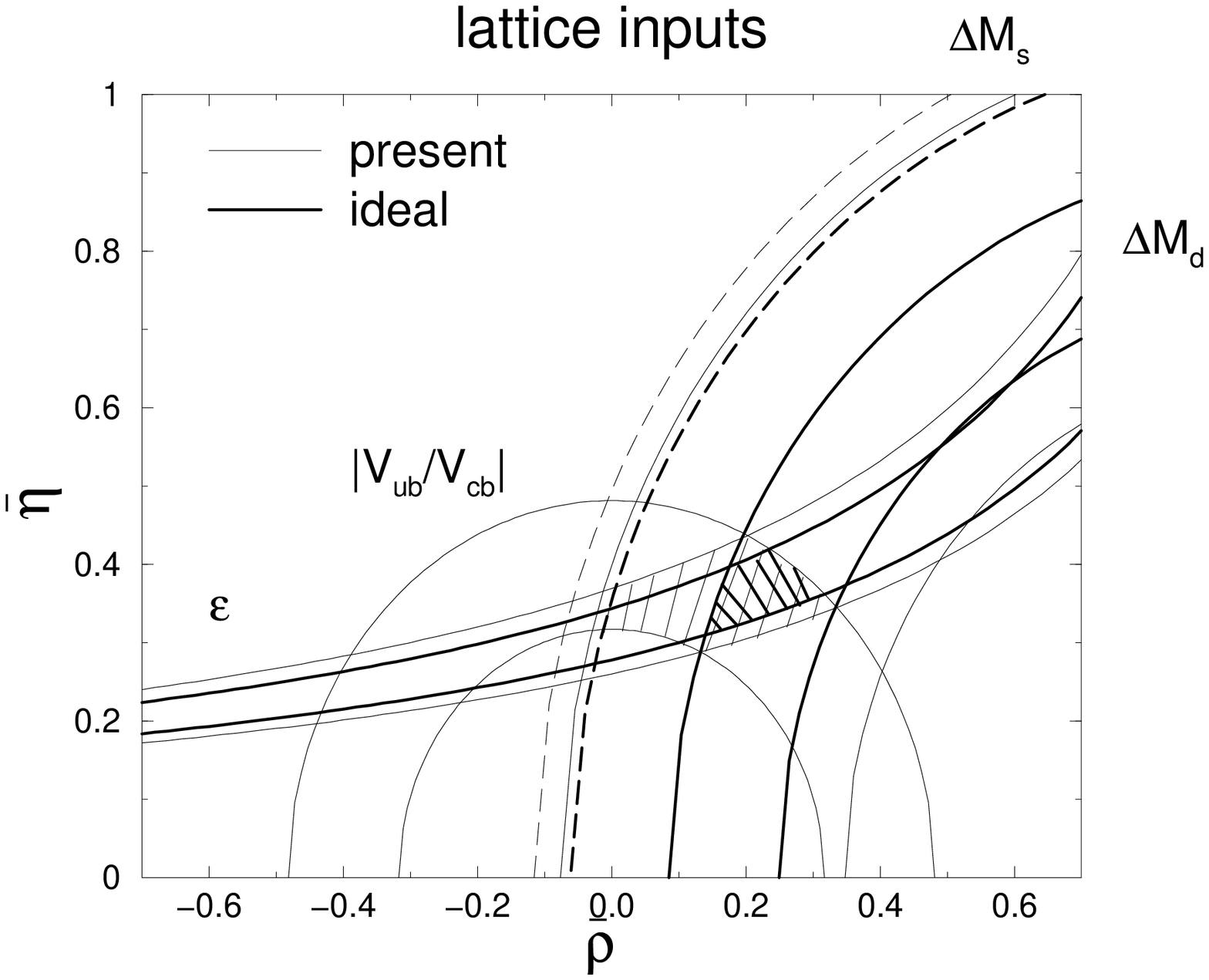, height = 5.7cm}
\caption{Constraints for CKM matrix in $\bar\rho-\bar\eta$ plane
from the standard input (left) and the lattice input (right)}
\label{fig:CKM}
\end{center}
\end{figure}

In Fig.~\ref{fig:CKM}
the plot on the left shows the output in the $\bar\rho-\bar\eta$ plane from the
standard input. The shaded region is allowed within 1 $\sigma$
of each constraint.
We take a flat distribution of each error in this analysis.
In the plot on the right the output from the lattice input is represented
by thin curves (denoted as ``present''). 
Due to the larger value of $\sqrt{\hat B_B}f_B$ than the
one in the standard input, more positive values of $\bar\rho$ are favored.
Thick curves (denoted as ``ideal'')
are constraints from lattice inputs without errors.
In both plots the constraint from $\Delta M_s$ is represented by dashed curves.

From these figures we conclude the following.
For a better determination of CKM matrix,
lattice calculations have to reduce errors in $\sqrt{\hat B_B} f_B$ and
precisely determine the central value of both $B_K$ and 
$\sqrt{\hat B_B} f_B$,
with less systematic uncertainties (in the continuum limit of full QCD).  
On the other hand, experiments need to 
reduce errors in $\vert V_{cb}\vert$ and $\vert V_{ub}\vert$,
where lattice calculations can also contribute,
and measure $\Delta M_s$ instead of a lower bound available at present.

\section{Summary}
\label{sec:end}
In this review we have presented the status of lattice QCD calculations of 
the quantities relevant for further understanding of the Standard Model 
and its limitations. 
The main results may be summarized as follows.
\begin{itemize}
\item In the continuum limit of full QCD with $N_f =2$, lattice QCD predicts
that $m_s^{\overline{MS}}(m_c)$ = 100 (15) MeV,
which is smaller than expected.
This small value of the strange quark mass tends to
increase the estimate for $\varepsilon'/\varepsilon$
in the standard analysis.
\item The continuum limit of quenched calculations for the $K_0- \bar K_0$ 
mixing parameter gives $\hat B_K = 0.87(6)$. Clearly the full QCD estimate
is called for.
\item Calculations in full QCD with $N_f=2$ for
the $B$ meson decay constant estimate $f_B = 210(30)$ MeV ,
which is found to be larger than previous quenched results.
\item The $B_0-\bar B_0$ mixing parameter $B_B(m_b) =$ 0.80(5) in the
static limit of the quenched approximation. The next step is
to estimate the heavy meson mass dependence.
\item Lattice methods for form factors of 
$D\rightarrow K^{(*)}$, $\pi$, $\rho$ decays
are now well established. There is a promising method for the calculation
of ${\cal F}_{B\rightarrow D^{(*)}} (1)$ to extract $\vert V_{cb}\vert$.
More works are still needed for $B\rightarrow \pi$ ($\rho$) decays.
\end{itemize}
As a conclusion we stress that
systematic investigations for all these quantities
in the continuum limit of full QCD will be the next target.

\bigskip
I thank Drs. P. Collins,
S. Hashimoto, H. Shanahan,
and A. Ukawa for informative communications and useful discussions.

\eject

\def\Discussion{
\setlength{\parskip}{0.3cm}\setlength{\parindent}{0.0cm}
     \bigskip\bigskip      {\Large {\bf Discussion}} \bigskip}
\def\speaker#1{{\bf #1:}\ }

\Discussion

\speaker{Masanori Yamauchi (KEK)}
Do the  errors quoted represent one standard deviation,  just
 like experimental statistical errors?

\speaker{Aoki}
  Yes, the quoted numbers generally represent one standard deviation of
statistical errors, while some of them also
include systematic errors.

\speaker{Jeff Richman (UCSB)}
I would like to comment that there is no fundamental obstacle to 
measuring the rate for
$B\rightarrow \rho \ell \nu$ at high $q^2$.  In fact, the experimental 
acceptance is good in
this region, and CLEO has now obtained a crude measurement of 
$d\Gamma/dq^2$ for
$B\rightarrow \rho\ell\nu$, which is limited only 
by statistics.  If we meet again in three years,
perhaps the lattice will have a solid prediction for 
the rate at high $q^2$ and
experimenters will have a solid measurement.

\speaker{Aoki}  I agree with you. I think that lattice theorists and 
experimentalists in this field should communicate with each other.

\speaker{George Hou (National Taiwan University)}
The strange quark mass 
$m_s$ not only enters $\varepsilon^\prime/\varepsilon$, but
 it also enters rare $B$ decays.  In
the CLEO fit discussed in Poling's talk 
that obtained a somewhat different value for $\gamma$ 
(or $\phi_2$) 
than usual, one also obtains a $m_s$ value 
consistent with the lower value that you 
have reported.

\speaker{Aoki}  That is an interesting information. I would like to see
the detailed analysis in the future.

\speaker{F. Constantini (Pisa University)}
It is true that the decrease of the strange $q$ mass from 130 MeV to 
100 improves the 
agreement between the lattice result and the experimental value
 of $\varepsilon^\prime/\varepsilon$.  However, to agree with the present
experimental values you need to go as low as 70 
MeV.  Moreover the dependence on the strange quark
 mass acts on the numerator of
the expression and not only on the denominator.

\speaker{Aoki} Since there exist ambiguities also in other parameters
such as $B_6$, $B_8$ and $\Lambda_{\bar{MS}}$,
we do not need the strange quark mass as low as 70 MeV.
The explicit strange quark mass dependence in the numerator of the expression
is included in my estimate. As mentioned in the talk, however,
the implicit strange quark mass dependence in $B_6$ and $B_8$ is not
considered here.

\speaker{Matthias Neubert (SLAC)}
In fact the quantity $\varepsilon^\prime/\varepsilon$ does not depend at 
all on the strange quark
mass.  It enters only by convention when you define $B_6$. 
 It seems to me you should
not use a lattice estimate of $m_s$ unless at the same time you 
use a lattice estimate
of $B_6$.

\speaker{Aoki} Rigorously speaking, I agree with you.
According to this criteria, however, I have to say
that there is no reliable theoretical estimate 
for $\varepsilon^\prime/\varepsilon$. 
Now it seems for me that only lattice calculations can give
a reliable theoretical estimate for $\varepsilon^\prime/\varepsilon$,
and, as I mentioned in my talk, serious efforts on such calculations have just 
started.

\speaker{B. F. L. Ward (University of Tennessee)}
Could you comment on the prospects for calculating the rate of 
$B\rightarrow \pi\pi$ on the lattice?

\speaker{Aoki}  
There are two main difficulties for calculating the rate of 
$B\rightarrow \pi\pi$ on the lattice. One is the problem 
of dealing with decays into 2 particles suck as $K\rightarrow\pi\pi$
decays, the other is the difficulty of putting a heavy quark on the lattice.
Therefore, no lattice calculation has been made for the
rate of $B\rightarrow \pi\pi$ so far.

\speaker{Helen Quinn (SLAC)}
When you put your  errors on lattice quantities for which you have only a 
quenched calculation
I believe that these errors include statistics and errors due to 
extrapolation 
but no estimate of the errors due
to quenching.  Is this correct, and if so how can one 
best estimate the additional uncertainties due to
the quenched approximation? 

\speaker{Aoki}   
Yes. In general the errors due to quenching are not included, and it is
difficult to estimate them unless the corresponding full QCD calculation
is really performed. One might naively think that the quenching errors
are 10--20\%. 
In some calculations, the quenching errors are partly estimated by the 
variation of the result under the change of experimental input parameters.
For example, the difference of the strange quark mass between
$K$-input and $\Phi$-input, which is about 20--25\%, may be interpreted as
the quenching errors. However this kind of the estimation is far
from being quantitative.


\end{document}

%% file: LP99macros.tex
\def\beq{\begin{equation}}
\def\eeq#1{\label{#1}\end{equation}}
\def\eeqn{\end{equation}}

\def\beqa{\begin{eqnarray}}
\def\eeqa#1{\label{#1}\end{eqnarray}}
\def\eeqan{\end{eqnarray}}

\def\NPB{ Nucl. Phys. B}
\def\PLB{Phys. Lett.  B}
\def\PRL{Phys. Rev. Lett.}
\def\PRD{Phys. Rev. D}

\let\bar=\overbar

\def\Dslash{\not{\hbox{\kern-4pt $D$}}}
\def\dslash{\not{\hbox{\kern-2pt $\del$}}}

\def\msb{{\bar{\ssstyle M \kern -1pt S}}}




%% file: lp99.bbl
\begin{thebibliography}{99}
\bibitem{KTeV}
Edward Blucher, 
talk at Lepton-Photon 99, 
in these proceedings.

\bibitem{Na48}
Giles Barr,
talk at Lepton-Photon 99, 
in these proceedings.

\bibitem{babar}
Jonathan Dorfan
talk at Lepton-Photon 99, 
in these proceedings.

\bibitem{belle}
Fumihiko Takasaki,
talk at Lepton-Photon 99, 
in these proceedings.

\bibitem{lat99}
S. Hashimoto
hep-lat/9909136.

\bibitem{lat98}
T. Draper,
\NPB (Proc. Suppl.)73, 43 (1999).

\bibitem{qCP-PACS}
CP-PACS Collaboration: S. Aoki {\it et al.}, 
hep-lat/9904012.

\bibitem{GF11}
F. Butler {\it et al.},
\PRL 70, 2849 (1993);

\bibitem{fCP-PACS}
CP-PACS Collaboration: A.Ali Khan {\it et ak.}, 
hep-lat/9909050.

\bibitem{wilson}
K.G. Wilson, in New Phenomena in Subnuclear Physics,
ed. by A. Zichichi (Plenum Press, New York, 1977).

\bibitem{BG}
T. Bhattacharya and R. Gupta,
\NPB (Proc. Suppl.) 63A-C, 95 (1998).

\bibitem{sw}
B. Sheikholeslami and R. Wohlert, \NPB 259, 527 (1985).

\bibitem{ks}
L. Susskind, \PRD 16, 3031 (1977).

\bibitem{SumRule}
L.Lellouch, E. de Rafael, J. Taron,
\PLB 414, 195 (1997);

\bibitem{NPZ}
S. Capitani, M. L\"uscher, R. Sommer, H. Wittig,
\NPB 544, 669 (1999).

\bibitem{jlqcd_wbk}
JLQCD Collaboration: S. Aoki {\it et al.},
\PRL 80, 1778 (1998)

\bibitem{lanl_wbk}
R. Gupta, T Bhattacharya, S.R. Sharpe,
\PRD 55, 4036 (1997).

\bibitem{ukqcd_cbk}
UKQCD Collaboration: L. Lellouch, C.-J.D. Lin,
\NPB (Proc. Suppl.) 73, 312 (1999),

\bibitem{rome_cbk}
L. Conti {\it et al.}, 
\PLB 421, 273(1998).

\bibitem{dwf}
D. Kaplan,
\PLB 288, 342 (1992);
Y. Shamir,
\NPB 409, 90 (1993);
V. Furuman, Y. Shamir,
\NPB 439, 54 (1995).

\bibitem{bs_dwbk}
T. Blum, A. Soni,
\PRL 79, 3595 (1997).

\bibitem{rbc_dwbk}
Riken-BNL-Columbia collaboration: T. Blum, A. Soni,
hep-lat/9909108.

\bibitem{jlqcd_ksbk}
JLQCD Collaboration: S. Aoki {\it et al.},
\PRL 80, 5271 (1998) .

\bibitem{sharpe}
S.R. Sharpe,
\NPB (Proc. Suppl.)53, 181 (1997).

\bibitem{KPV}
G. Kilcup {\it et al.},
\NPB (Proc. Suppl.)53, 345 (1997).

\bibitem{buras99}
A. J. Buras,
hep-ph/9905437.

\bibitem{DGGM}
A. Donini {\it et al.},
hep-lat/9910017.

\bibitem{MT}
L. Maiani, M. Testa,
\PLB 245, 585 (1990).

\bibitem{bdspw}
C. Bernard {\it et al.}, 
\PRD 32, 2343 (1985)

\bibitem{kp_kpp}
D. Pekurovsky, G. Kilcup,
hep-lat/9812019.

\bibitem{rbc_kpp}
Riken-BNL-Columbia collaboration: T. Blum {\it et al.}, 
hep-lat/9908025.

\bibitem{kgs97}
G. Kilcup, R. Gupta, S.R. Sharpe,
\PRD 57, 1654 (1998).

\bibitem{jlqcd_a2}
JLQCD Collaboration: S. Aoki {\it et al.},
\PRD 58, 054503 (1998).

\bibitem{fnal}
A.X. El-Khadra, A.S. Kronfeld, P.B. Mackenzie,
\PRD 55, 3933 (1997).

\bibitem{nrqcd}
B.A. Thacker, G.P. Lepage,
\PRD 43, 196 (1991).

\bibitem{wittig}
H. Wittig,
Int. J. Mod. Phys. A12, 4477 (1997),
hep-lat/9705034.

\bibitem{jlqcd_nrfb}
JLQCD collaboration: K.-I. Ishikawa {\it et al.},
hep-lat/9905036.

\bibitem{glok_nrfb}
A. Ali Khan {\it et al.},
\PLB 427, 132 (1998).

\bibitem{cppacs_nrfb}
CP-PACS collaboration: A. Ali Khan {\it et al.},
hep-lat/9911039.

\bibitem{milc_fnfb}
C. Bernard {\it et al.},
\PRL 81, 4812 (1998).

\bibitem{fnal_fnfb}
A.X. El-Khadra {\it et al.},
\PRD 58, 014506 (1998).

\bibitem{jlqcd_fnfb}
JLQCD collaboration: S. Aoki {\it et al.},
\PRL 80, 5711 (1998).

\bibitem{glok_nrfb_full}
S. Collins {\it et al.},
hep-lat/9901001.

\bibitem{milc_fnfb_full}
S. Gottlieb {\it et al.},
hep-lat/9909121.

\bibitem{cppacs_fnfb_full}
CP-PACS collaboration: A. Ali Khan {\it et al.},
hep-lat/9909052.

\bibitem{LL}
UKQCD Collaboration: L. Lellouch, C.-J.D.Lin, 
talk at Heavy Flavours 8.

\bibitem{kent_bb}
J. Christensen, T. Draper, C. MacNeile,
\PRD 56, 6993 (1997).

\bibitem{gm_bb}
V. Gimenez and G. Martinelli,
\PLB 398, 135 (1997).

\bibitem{gr_bb}
V. Gimenez and J. Reyes,
\NPB 545, 576 (1999).

\bibitem{ukqcd_bb}
UKQCD Collaboration: A.K. Ewing {\it et al.},
\PRD 54, 3526 (1996).

\bibitem{delphi}
DELPHI Collaboration: paper submitted to the HEP'99 Conference.

\bibitem{hiroshima_bb}
S. Hashimoto {\it et al.},
hep-lat/9903002.

\bibitem{bernard_bb}
C. Bernald, T. Draper, G. Hockney, A. Soni,
\PRD 38, 3540 (1988).

\bibitem{elc_bb}
A. Abada {\it et al.},
\NPB  376, 172 (1992).

\bibitem{bs_bb}
A. Soni,
\NPB (Proc. Suppl.) 47, 43 (1996).

\bibitem{ukqcd_clbb}
UKQCD Collaboration: L. Lellouch, C.-J. D. Lin,
%
\NPB (Proc. Suppl.) 73, 357 (1999).

\bibitem{ukqcd_d2k}
UKQCD collaboration: K.C. Bowler {\it et al.},
\PRD 51, 4905 (1995).

\bibitem{FS}
J.M. Flynn, C.T. Sachrajda, in Heavy Flavours (2nd ed.),
ed. by A.J. Buras and M. Linder (World Scientific, Singapore),
hep-lat/9710057.

\bibitem{fnal_b2d}
S. Hashimoto {\it et al.}, 
hep-ph/9906376.

\bibitem{SR_b2d}
I.I. Bigi,
hep-ph/9906376.

\bibitem{ukqcd_b2d}
UKQCD collaboration: G. Douglas,
hep-lat/9909126.

\bibitem{cleo_b2r}
CLEO collaboration: B.H. Behrens {\it et al.},
hep-ex/9905056.

\bibitem{ukqcd_b2r}
UKQCD collaboration: J.M. Flynn {\it et al.},
\NPB 447, 425 (1996).

\bibitem{lat97}
T. Onogi,
\NPB (Proc. Suppl.) 63, 59 (1998).

\bibitem{ukqcd_b2p}
UKQCD collaboration: C.M. Maynard {\it et al.},
hep-lat/9909100.

\end{thebibliography}
